\documentclass[journal]{IEEEtran}
\ifCLASSINFOpdf
  \usepackage[pdftex]{graphicx}
   \usepackage{epstopdf}
\else
\fi
\usepackage{amsmath} 
\usepackage{amsfonts} 
 \usepackage{float}
 \usepackage{epstopdf}
\usepackage{mathrsfs}
\usepackage{pifont}
\usepackage{bbding}
\usepackage{amsmath}
\usepackage[utf8]{inputenc}  
\usepackage[T1]{fontenc}
\usepackage{multirow}
\usepackage{tipa}
\usepackage{algorithm}
\usepackage{algorithmic}
\usepackage{amsmath,epsfig}
\usepackage{amssymb}
\usepackage{amsfonts}
\usepackage{epic}
\usepackage{graphicx}
\usepackage{curves}
\usepackage{subfigure}
\usepackage[justification=centering]{caption}
\usepackage{bm}
\usepackage{color}
\ifCLASSINFOpdf
\else
\fi

\begin{document}
\newtheorem{theorem}{Theorem}
\newtheorem{proposition}{Proposition}
\newtheorem{definition}{Definition}
\newtheorem{lemma}{Lemma}
\newtheorem{corollary}{Corollary}
\newtheorem{remark}{Remark}
\newtheorem{construction}{Construction}

\newcommand{\supp}{\mathop{\rm supp}}
\newcommand{\sinc}{\mathop{\rm sinc}}
\newcommand{\spann}{\mathop{\rm span}}
\newcommand{\essinf}{\mathop{\rm ess\,inf}}
\newcommand{\esssup}{\mathop{\rm ess\,sup}}
\newcommand{\Lip}{\rm Lip}
\newcommand{\sign}{\mathop{\rm sign}}
\newcommand{\osc}{\mathop{\rm osc}}
\newcommand{\R}{{\mathbb{R}}}
\newcommand{\Z}{{\mathbb{Z}}}
\newcommand{\C}{{\mathbb{C}}}
%

\title{ Efficient Spectral Efficiency Maximization Design for IRS-aided MIMO Systems }
\author{{Fuying~Li, Yajun Wang,   Zhuxian~Lian,~\IEEEmembership{Member,~IEEE},  and Wen Chen,~\IEEEmembership{Senior Member,~IEEE}}
\thanks{Fuying~Li, Yajun~Wang, and Zhuxian~Lian are with the Department of Electronic Engineering,
Jiangsu  University of Science and Technology, Zhenjiang, 212003, China (e-mail: leeonzheearth@163.com;  wangyj1859@just.edu.cn;  zhuxianlian@just.edu.cn).}
\thanks{ Wen Chen are with the Department of Electronic Engineering, Shanghai Jiao Tong University, Shanghai, 200240, China,
 (e-mail:  wenchen@sjtu.edu.cn).}
\thanks{This work was supported in part by the National
Key R\&D Program of China 2023YFB2905000; in part by the National
Key Project 2020YFB1807700; in part by FDCT
under Grant 0119/2020/A3; in part by GDST under Grant 2020B1212030003;
in part by STDF, Macau, under Grant 0036/2019/A1; and in part by
NSFC under Grant 62001194, 61872184.}}

%

\maketitle
\begin{abstract}
Driven by the growing demand for higher spectral efficiency in wireless communications, intelligent reflecting surfaces (IRS) have attracted considerable attention for their ability to dynamically reconfigure the propagation environment. This work addresses the spectral efficiency maximization problem in IRS-assisted multiple-input multiple-output (MIMO) systems, which involves the joint optimization of the transmit precoding matrix and the IRS phase shift configuration. This problem is inherently challenging due to its non-convex nature. To tackle it effectively, we introduce a computationally efficient algorithm, termed ADMM-APG, which integrates the alternating direction method of multipliers (ADMM) with the accelerated projected gradient (APG) method. The proposed framework decomposes the original problem into tractable subproblems, each admitting a closed-form solution while maintaining low computational complexity.
~Simulation results demonstrate that the ADMM-APG algorithm consistently surpasses existing benchmark methods in terms of spectral efficiency and computational complexity, achieving significant performance gains across a range of system configurations.
\end{abstract}



%
\IEEEpeerreviewmaketitle
\section{Introduction}\label{sec:1}
The convergence of mobile internet, IoT, and artificial intelligence has accelerated deployment of data-intensive applications-from immersive media to industrial automation-imposing unprecedented requirements for ultra-high speed, minimal latency, and extreme reliability in communication networks. This transformation has driven exponential growth in global wireless data traffic, intensifying pressure on finite spectrum resources and pushing conventional systems toward fundamental capacity limits. Simultaneously, the dense deployment of infrastructure needed to support these services has resulted in unsustainable energy consumption patterns, where base station (BS) circuit losses and-increasingly critical-cooling system overhead dominate as constraints on network energy efficiency \cite{IEEEconf:1}.

These challenges are exacerbated by complex propagation environments, where urban high-rises, intricate indoor layouts, and specialized scenarios introduce severe attenuation and coverage gaps. Conventional solutions such as power amplification or relay deployment tend to compound energy inefficiency while raising system complexity and cost \cite{IEEEconf:2}. In response, intelligent reflecting surface (IRS) technology has emerged as a transformative approach. Constructed from programmable metamaterial elements, IRS dynamically shapes electromagnetic waves by electronically controlling the phase and amplitude of incident signals \cite{IEEEconf:3, IEEEconf:4}. This enables precise wavefront manipulation, focusing energy toward intended users or steering beams around obstacles to extend coverage and enhance link reliability \cite{IEEEconf:5}. Unlike traditional approaches, IRS-assisted transmission mitigates path loss and fading more effectively-especially in non-line-of-sight settings-without requiring additional power-intensive amplification \cite{IEEEconf:6, IEEEconf:7}. Through optimized beamforming, IRS significantly improves spectral efficiency, data rates, and transmission robustness, enabling reliable communications over broader areas \cite{IEEEconf:8}-\cite{IEEEconf:11}.

IRS technology continues to attract extensive research attention across wireless domains. Early contributions \cite{IEEEconf:12}-\cite{IEEEconf:14} established foundational designs for single-user setups, followed by extensions to multi-user settings addressing interference management and resource allocation \cite{IEEEconf:15}-\cite{IEEEconf:17}. More recently, integration with massive multiple-input multiple-output (MIMO) systems has become an active frontier, given its potential for substantial performance gains.

A central challenge in IRS-aided MIMO architectures lies in the coupled optimization of active precoding at the BS, receive processing at users, and passive reflection at the IRS \cite{IEEEconf:18}-\cite{IEEEconf:19}. These joint optimization problems are inherently non-convex, and their computational complexity grows prohibitively with the number of IRS elements-posing a major obstacle to practical large-scale implementation \cite{IEEEconf:20}, \cite{IEEEconf:21}.

Existing research has extensively investigated weighted sum-rate (WSR) maximization through joint optimization of active and passive beamforming configurations \cite{IEEEconf:22, IEEEconf:23}. Parallel developments have focused on enhancing both energy and spectral efficiency in IRS-assisted wireless and unmanned aerial vehicle (UAV) networks \cite{IEEEconf:24, IEEEconf:25}, with complementary investigations addressing fairness-oriented communication through max-min signal-to-interference-plus-noise ratio (SINR) optimization frameworks \cite{IEEEconf:26}. Further extending this direction, \cite{IEEEconf:27} demonstrates how coordinated optimization of active array beamforming at access points and passive phase-shift beamforming at IRSs can minimize total transmit power while maintaining satisfactory SINR levels for all users. The scope of joint precoding has been further expanded to encompass heterogeneous network architectures incorporating multi-BS, multi-IRS, multi-user, and multi-carrier configurations \cite{IEEEconf:28}.

Emerging research directions include IRS integration with millimeter-wave hybrid beamforming \cite{IEEEconf:29}, \cite{IEEEconf:30}, resource allocation in orthogonal frequency division multiplex (OFDM) systems \cite{IEEEconf:31}, radar-communication coexistence \cite{IEEEconf:32}, mobile edge computing with binary offloading \cite{IEEEconf:33}, and integration with movable antennas \cite{IEEEconf:34}. Multi-IRS cooperation has been investigated through double-IRS beamforming designs \cite{IEEEconf:35} and associated channel modeling \cite{IEEEconf:36}, while cell-free massive MIMO represents another promising application area \cite{IEEEconf:37}-\cite{IEEEconf:39}.

This paper investigates an IRS-assisted MIMO communication system, where a multi-antenna BS communicates with a multi-antenna user through an IRS. The system aims to maximize spectral efficiency under practical constraints imposed by the BS precoder and IRS phase shift configuration. To achieve this goal, we jointly optimize the active transmit precoding matrix at the BS and the passive reflection matrix at the IRS, establishing a comprehensive framework for enhancing system performance through coordinated active and passive beamforming design.

Maximizing spectral efficiency in IRS-aided MIMO systems has been addressed through several methodologies. The sum-path gain maximization (SPGM) approach \cite{IEEEconf:40} employs the alternating direction method of multipliers (ADMM) to jointly optimize transmitter precoding and IRS phase shifts, thereby improving overall path gain and spectral efficiency. However, its computational complexity increases cubically with the number of IRS phase shifts, which limits practical deployment. To mitigate this issue, a linearized ADMM (LADMM) variant [41] introduces a linear approximation strategy that not only alleviates computational load but also enhances spectral efficiency. Another method, the dimension-wise sinusoidal maximization (DSM) algorithm \cite{IEEEconf:42}, exploits the sinusoidal characteristics of individual reflecting element phase shifts and adopts sequential alternating optimization to maximize sum capacity with reduced complexity. Despite this advantage, DSM's element-wise optimization framework hampers its performance in large-scale IRS configurations. Further advancements include a Riemannian gradient descent network that maintains low complexity while pursuing the same objective as SPGM \cite{IEEEconf:43}. Alternating optimization (AO) \cite{IEEEconf:44}, though straightforward to implement for improving data rates in IRS-assisted MIMO systems, tends to converge slowly and incurs high computational costs, especially with large IRS arrays. In contrast, the projected gradient method (PGM) \cite{IEEEconf:45} achieves comparable rate performance to AO with fewer iterations and lower complexity, resulting in significantly improved operational efficiency.

While existing approaches have advanced spectral efficiency maximization in IRS-assisted MIMO systems, they generally converge to suboptimal solutions or incur high computational complexity, leaving substantial room for performance improvement. Motivated by these limitations, we revisit the spectral efficiency maximization problem with the aim of developing a computationally efficient algorithm that overcomes the constraints of prior methods. The main contributions of this work are summarized as follows:

$\bullet$  We propose a novel ADMM-APG algorithm that integrates the accelerated projected gradient (APG) method into the ADMM framework. This hybrid approach decomposes the original problem into three tractable subproblems: precoding matrix, auxiliary matrix, and phase-shift matrix, enabling efficient and stable optimization.

$\bullet$ This approach yields a closed-form solution for every subproblem, eliminating iterative optimization steps, streamlining the overall optimization process, and substantially reducing computational overhead.

$\bullet$ Theoretical analysis confirms that the computational complexity of ADMM-APG is competitive with state-of-the-art methods. Simulations further demonstrate that the proposed algorithm achieves higher spectral efficiency and faster convergence compared to existing benchmarks, validating its effectiveness and practical advantage.

Notation: Vectors and matrices are denoted by boldface lower- and upper-case letters, respectively. The space of $a \times b$ complex matrices is represented by $\mathbb{C}^{a \times b}$. The transpose, complex conjugate, and Hermitian transpose operators are denoted by $(\cdot)^T$, $(\cdot)^*$, and $(\cdot)^H$, respectively. The natural logarithm of $x$ is written as $\ln(x)$. For notational simplicity, $||\cdot||$ denotes the Euclidean norm for vectors and the Frobenius norm for matrices. The operator $\text{diag}(\mathbf{x})$ generates a square diagonal matrix with the elements of $\mathbf{x}$ on its main diagonal, and $|x|$ gives the absolute value of $x$. The $l$-th entry of vector $\mathbf{x}$ is denoted by $x_l$. The trace and determinant of matrix $\mathbf{X}$ are written as $\text{Tr}(\mathbf{X})$ and $\det(\mathbf{X})$, respectively. $\mathbf{I}_{K}$ indicates a $K\times K$  identity matrix.  
~The random vector $\mathbf{d}$ follows a circularly symmetric complex Gaussian distribution, $\mathbf{d} \sim \mathcal{CN}(\mathbf{0}, \mathbf{\Gamma})$, with mean $\mathbf{0}$ and covariance matrix $\mathbf{\Gamma}$. The gradient of function $f$ with respect to $\mathbf{X}^* \in \mathbb{C}^{m \times n}$ is denoted by $\nabla_{\mathbf{X}} f(\cdot)$.
~The operator $\textrm{vec}_{d}(\mathbf{X})$ forms a vector from the diagonal elements of $\mathbf{X}$, while $\text{vec}(\mathbf{X})$ vectorizes $\mathbf{X}$ by stacking its columns. Finally, $A_{ik}$ refers to the element in the $i$-th row and $k$-th column of matrix $\mathbf{A}$.


\section{System Model and Problem Formulation }\label{sec:2}
We examine a MIMO wireless communication system comprising a base station (BS) equipped with $M_t$ transmit antennas and a receiver with $M_r$ receive antennas. An IRS with $M_i$ passive elements is deployed to improve the communication link. The system operates over a narrowband frequency-flat channel, and full channel state information (CSI) is assumed to be available at a centralized controller. Each IRS element is assumed ideal and capable of independently adjusting both the phase shift and reflection angle of incident waves. Furthermore, due to significant path loss, signals undergoing multiple reflections at the IRS are considered negligible and are therefore disregarded in the model.

Let $\mathbf{H}_2 \in \mathbb{C}^{M_r \times M_t}$ denote the direct channel between the transmitter and the receiver, $\mathbf{H}_m \in \mathbb{C}^{M_i \times M_t}$ represent the channel from the transmitter to the IRS, and $\mathbf{H}_1 \in \mathbb{C}^{M_r \times M_i}$ correspond to the channel from the IRS to the receiver. At the transmitter, the data symbol vector $\mathbf{d} \in \mathbb{C}^{M_s \times 1}$ (where $M_s$ is the number of data streams), distributed as $\mathbf{d} \sim \mathcal{CN}(\mathbf{0}, \mathbf{I}_{M_s})$, is precoded by a linear precoding matrix $\mathbf{G} \in \mathbb{C}^{M_t \times M_s}$. The precoded signal is then simultaneously transmitted to both the receiver and the IRS.

\begin{figure}
	\centering
	\includegraphics[width=3in,angle=0]{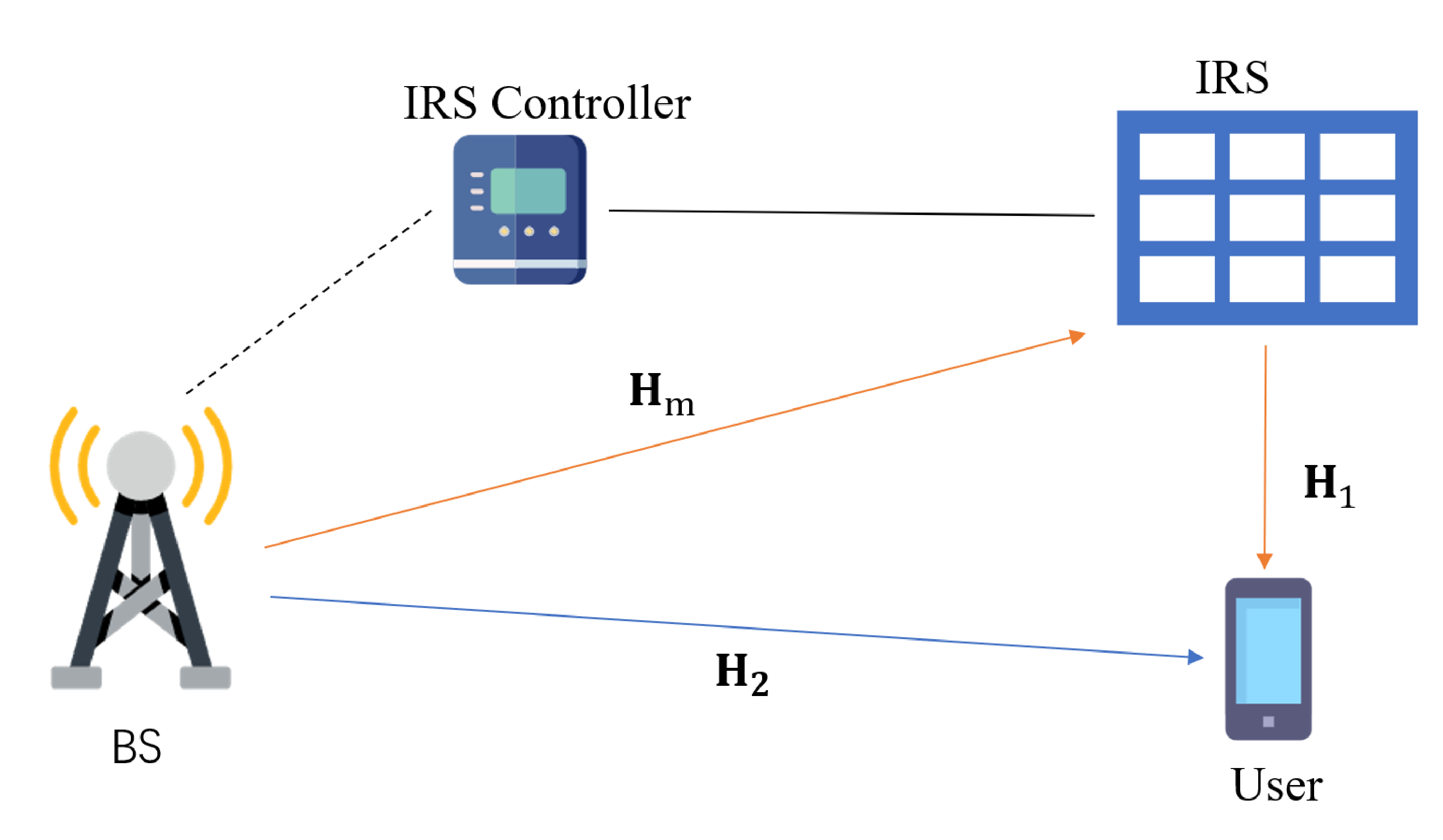}
	\caption{System model.}
	\label{fig1}
\end{figure}

 The reflection matrix of the IRS is defined as $\boldsymbol{\Phi} = \mathrm{diag}(\boldsymbol{\theta})\in\mathbb{C}^{M_i \times M_i}$, where $\boldsymbol{\theta} = \left[ e^{j\varphi_1}, e^{j\varphi_2}, \dots, e^{j\varphi_{M_{i}}} \right]^T$  represents the phase adjustments of each reflecting element, and
 $\varphi_i \in [0, 2\pi], i = 1, 2, \dots, M_i$ represents the phase shift angle of each reflecting element.

Based on above assumption, the total received signal at the user can be expressed as \cite{IEEEconf:32}:
\begin{equation}\label{eq1}
	\mathbf{y} = \sqrt{\frac{P}{M_s}} \left( \mathbf{H}_1 \boldsymbol{\Phi} \mathbf{H}_m  + \mathbf{H}_2\right) \mathbf{G} \mathbf{d}  + \mathbf{n},
\end{equation}
where, $P$ denotes the total transmit power at the transmitter, and $\mathbf{n}\sim\mathcal{C}\mathcal{N}(\mathbf{0},\sigma_{n}^{2}{{\mathbf{I}}_{{{M}_{s}}}})$ represents the  additive white Gaussian noise. The term $\mathbf{H}_{1}\mathbf{\Phi }\mathbf{H}_{m}\mathbf{G}\mathbf{d}$ corresponds to the signal  through the BS-IRS-user  link, while the term $\mathbf{H}_{2}\mathbf{G}\mathbf{d}$ corresponds to the signal component transmitted through the direct link. Together, these two components form the useful signal portion in the received signal.

The equivalent channel of the entire system can be expressed as
$\mathbf{H} = \mathbf{H}_1 \boldsymbol{\Phi} \mathbf{H}_m + \mathbf{H}_2$.
To harness the full spatial multiplexing potential of the channel and maximize the overall spectral efficiency, we consider the number of effective spatial streams
 ${|| {\bf{G}} ||_F^{2}=M_s=rank(\mathbf{H})}$. 
Consequently, the problem of maximizing the spectral efficiency in the given system can be formulated as the following optimization.
\begin{equation}\label{eq2}
	\begin{aligned}
		\max_{\mathbf{G}, \boldsymbol{\theta}} \quad &R=\log_2 \det \left(\mathbf{I}_{M_r} + \frac{P}{\sigma_n^2 M_s} \mathbf{H} \mathbf{G} \mathbf{G}^H \mathbf{H}^H \right) \\
		\text{s.t.} \quad &\mathbf{H} = \mathbf{H}_1 \boldsymbol{\Phi} \mathbf{H}_m + \mathbf{H}_2, \\
		&\ ||\mathbf{G}||_F^2 = M_s, \, \boldsymbol{\Phi} = \mathrm{diag}(\boldsymbol{\theta}), \\
		&|{\theta_{n}}| = 1, \, n = 1, 2, \ldots, M_i.
	\end{aligned}
\end{equation}
Given the non-convex nature of both the objective function and the unit-modulus constraint $|{\theta_{n}}|=1$, the optimization problem (2) becomes inherently non-convex and challenging to solve directly. To tackle this issue, we incorporate an APG approach within the ADMM framework, enabling efficient solution of the spectral efficiency maximization problem in the following section.

\section{Joint Beamforming via ADMM-APG }
In this section, we first employ the ADMM to address problem (2), subsequently incorporating the APG technique to optimize the reflection matrix $\boldsymbol{\Phi}$. Both methods are well-established in optimization theory, with extensive applications spanning diverse convex and non-convex problems [46], [47]. For detailed discussions of their theoretical foundations and implementation aspects, we refer readers to the comprehensive treatments in [46], [47].

\subsection{ADMM Framework}

Let $\mathbf{Y} = \mathbf{I}_{M_r} + C \mathbf{H G G}^\mathrm{H} \mathbf{H}^\mathrm{H}$, where $C=\frac{P}{\sigma_n^2 M_s}$. The problem (2) is transformed into the following optimization.
\begin{equation}\label{eq3}
	\begin{aligned}
		\min_{\mathbf{G}, \boldsymbol{\theta},\mathbf{Y}} \quad &-\log_{2} \det(\mathbf{Y})\\
		\text{s.t.} \quad & \mathbf{Y} = \mathbf{I}_{M_r} + C \mathbf{H G G}^\mathrm{H} \mathbf{H}^\mathrm{H},\\
		&\ ||\mathbf{G}||_F^2 = M_s, \, \boldsymbol{\Phi} = \mathrm{diag}(\boldsymbol{\theta}), \\
		&|{\theta}_n| = 1, \, n = 1, 2, \ldots, M_i
	\end{aligned}
\end{equation}

By the linear transformation, the precoding matrix $\mathbf{G}$ and equivalent channel $\mathbf{H}$ (or $\boldsymbol{\Phi}$)  is decoupled from the log-determinant term.
~The augmented Lagrangian function of (3) is given by
\begin{equation}\label{eq4}
\begin{aligned}
	\mathcal{L}_\rho(\mathbf{G}, \mathbf{Y}, \boldsymbol{\theta}, \mathbf{Z}) &= -\ln \det (\mathbf{Y})\\
&+ \frac{\rho}{2} \bigl\lVert \mathbf{Y} - \mathbf{I}_{M_r} - \mathbf{C H G G}^\mathrm{H} \mathbf{H}^\mathrm{H} + \mathbf{Z} \bigr\rVert_F^2,
\end{aligned}
\end{equation}
where $\mathbf{Z}$ is scaled dual matrix.

ADMM consists of the following iterations. 
\begin{subequations}\label{eq5}
	\begin{align}
		&\mathbf{G}^{k+1} = \underset{|| \mathbf{G} ||_F^{2} = M_s}{\arg\min} \, \mathcal{L}_{\rho} (\mathbf{G}, \mathbf{Y}^k, \boldsymbol{\theta}^k, \mathbf{Z}^k)\label{eq5a}\\
		&\mathbf{Y}^{k+1} = \arg\min \, \mathcal{L}_{\rho} (\mathbf{G}^{k+1}, \mathbf{Y}, \boldsymbol{\theta}^k, \mathbf{Z}^k)\label{eq5b}\\
		&\boldsymbol{\theta}^{k+1} = \underset{|{\theta}_n| = 1}{\arg\min} L_{\rho} \left( \mathbf{G}^{k+1}, \mathbf{Y}^{k+1}, \boldsymbol{\theta}, \mathbf{Z}^k \right)\label{eq5c}\\
		&\mathbf{Z}^{k+1} = \mathbf{Z}^k + \mathbf{Y}^{k+1} - \mathbf{I}_{M_r} - C \mathbf{H}^{k+1} \mathbf{G}^{k+1} (\mathbf{G}^{k+1})^\mathrm{H} (\mathbf{H}^{k+1})^\mathrm{H}\label{eq5d}
	\end{align}
\end{subequations}
where $k$ denotes the number of iterations. 

\subsection{Fix $\mathbf{Y}$, $\boldsymbol{\theta}$ and $\mathbf{Z}$ and Solve $\mathbf{G}$}
\
The subproblem (5a) for $\mathbf{G}$  is equivalent to the following optimization.
\begin{equation}
	\begin{aligned}\label{eq6}
		\min \ &|| \mathbf{Y}^k - \mathbf{I}_{M_r} - C \mathbf{H G} \mathbf{G}^H \mathbf{H}^\mathrm{H} + \mathbf{Z}^k ||_F^2 \\
		\text{s.t.} \quad& || \mathbf{G} ||_F^2 = M_s
	\end{aligned}
\end{equation}

Through the truncated singular value decomposition (SVD) of the effective channel $\mathbf{H}$, we obtain $\mathbf{H} = \mathbf{U} \boldsymbol{\Lambda} \mathbf{V}^H$, where $\mathbf{U} \in \mathbb{C}^{M_r \times M_s}$ and $\mathbf{V} \in \mathbb{C}^{M_t \times M_s}$ are unitary matrices. The matrix $\boldsymbol{\Lambda} \in \mathbb{C}^{M_s \times M_s}$ is a diagonal matrix consisting of $M_s$ singular values of $\mathbf{H}$ arranged in descending order.

Based on the SVD of the effective channel $\mathbf{H}$, the corresponding optimal precoding matrix is given by
\begin{equation}\label{eq7}
\mathbf{G}^{k+1} = \mathbf{V} \boldsymbol{A}^{\frac{1}{2}},
\end{equation}
where $\boldsymbol{A}=\text{diag}\{p_1,  p_2, \cdots,  p_{M_s}\}$ is  the water-filling power allocation matrix. Each element $p_j \geq 0$ corresponds to the power allocated to the $j$-th data stream  satisfying the constraint $\sum_{j=1}^{M_s} p_j = M_s$. The values of $p_j$ can be efficiently determined via the water-filling algorithm.

\subsection{Fix $\mathbf{G}$, $\boldsymbol{\theta}$ and $\mathbf{Z}$ and Solve $\mathbf{Y}$}
\
The subproblem (5b) requires solving the following unconstrained optimization.
\begin{equation}
	\begin{aligned}\label{eq8}
		\min\ &-\ln \det (\mathbf{Y}) \\
&+ \frac{\rho}{2} || \mathbf{Y} - \mathbf{I}_{M_r} - C \mathbf{H G}^{k+1} (\mathbf{G}^{k+1})^\mathrm{H} \mathbf{H}^\mathrm{H} + \mathbf{Z}^k ||_F^2
\end{aligned}
\end{equation}

Substituting $\mathbf{G}^{k+1} = \mathbf{V} \boldsymbol{A}^{\frac{1}{2}}$ into (7), we reformulate the optimization as follows.
\begin{equation}\label{eq9}
		\min\ -\ln \det (\mathbf{Y}) + \frac{\rho}{2} || \mathbf{Y} - \mathbf{I}_{M_r} - C \mathbf{U} \boldsymbol{\Lambda} \boldsymbol{A} \boldsymbol{\Lambda} \mathbf{U}^\mathrm{H} + \mathbf{Z}^k ||_F^2
\end{equation}

Solving for $\mathbf{Y}$ presents a significant challenge. To address this, we devise an efficient approach that leads to a closed-form solution, expressed as
\begin{equation}\label{eq10}
\mathbf{Y}^{k+1} = \mathbf{U}_1 \widetilde{\mathbf{Y}} \mathbf{U}_1^H,
\end{equation}
where $\mathbf{U}_1$ is obtained from the eigenvalue decomposition (EVD) of $\mathbf{I}_{M_r} + C \mathbf{U}_1 \boldsymbol{\Lambda} \boldsymbol{A} \boldsymbol{\Lambda} \mathbf{U}_1^H - \mathbf{Z}^k$, and $\widetilde{\mathbf{Y}}$  is a diagonal matrix whose diagonal entries are obtained by solving a set of quadratic equations. The complete derivation of this expression is detailed in Appendix A.
\subsection{Fix $\mathbf{G}$, $\mathbf{Y}$  and $\mathbf{Z}$, and Solve $\boldsymbol{\theta}$  by APG}
\
The subproblem (5c) can be recast as the following optimization.
\begin{equation}
	\begin{aligned}\label{eq11}
		\min \ &g(\boldsymbol{\theta})=|| \mathbf{Y}^{k+1} - \mathbf{I}_{M_r} - C \mathbf{H}\mathbf{G}^{k+1} (\mathbf{G}^{k+1})^H \mathbf{H}^\mathrm{H} + \mathbf{Z}^k ||_F^2 \\
		\text{s.t.} \quad& |{\theta_{n}}| = 1, \, n = 1, 2, \ldots, M_i.
	\end{aligned}
\end{equation}	
where $\mathbf{H}$ is  a function  of the parameter vector $\boldsymbol{\theta}$.

Owing to the unit-modulus constraints present in (11), the resulting optimization problem becomes non-convex and challenging to solve. To tackle it efficiently, we adopt the APG method. The iterative steps of the APG algorithm are outlined below.
 \begin{equation}\label{eq12}
 \begin{aligned}
	&\boldsymbol{\theta}^{k+1} = \text{Proj}_{|{\theta}_n|=1} \left( \boldsymbol{\omega}^k - \frac{1}{\tau^k} \nabla_{\boldsymbol{\theta}} g(\boldsymbol{\theta}^k) \right),\\
    &\boldsymbol{\omega}^k = \boldsymbol{\theta}^k + t_k \left( \boldsymbol{\theta}^k - \boldsymbol{\theta}^{k-1} \right)
\end{aligned}
\end{equation}
where $\text{Proj}$ denotes the projection operator, and $\tau^k>0$ and $t_k>0$ are  step sizes.

The step size $t_k$ is updated according to the following rule.
 \begin{equation}\label{eq13}
 t_k = \frac{d_k - 1}{d_k}, d_k = \frac{1 + \sqrt{1 + 4d_{k-1}^2}}{2}, d_0=0.
 \end{equation}

 Let $\mathbf{E} = \mathbf{Y}^{k+1} - \mathbf{I}_{M_r} - C \mathbf{H} \mathbf{G}^{k+1} (\mathbf{G}^{k+1})^\mathrm{H} \mathbf{H}^\mathrm{H} + \mathbf{Z}^k$, the gradient $\nabla_{\boldsymbol{\theta}} g(\boldsymbol{\theta})$ is given by
\begin{equation}\label{eq14}
\nabla_{\boldsymbol{\theta}} g(\boldsymbol{\theta}) = -2C\mathrm{vec}_{d} \left[ \mathbf{H}_1^\mathrm{H} \mathbf{E} \mathbf{H} \mathbf{G}^{k+1} (\mathbf{G}^{k+1})^\mathrm{H} \mathbf{H}_m^\mathrm{H} \right].
 \end{equation}
where $\mathrm{vec}_{d}(\cdot)$ extracts and vectorizes the diagonal entries of a matrix.

The detailed derivation of the gradient expression $\nabla_{\boldsymbol{\theta}} g(\boldsymbol{\theta})$ can be found in Appendix B.

Let $\boldsymbol{\xi}^{k}=\boldsymbol{\omega}^k - \frac{1}{\tau^k} \nabla_{\boldsymbol{\theta}} g(\boldsymbol{\theta}^k)$, since the phase shift elements must satisfy the unit modulus constraint, the projection step in (12) is applied element-wise as follows.
\begin{equation}\label{eq15}
	{\theta}_n^{k+1} =
	\begin{cases}
		\dfrac{{\xi}_n^k}{|{\xi}_n^k|} & ({\xi}_n^k \neq 0) \\[6pt]
		e^{j\varphi}, \, \varphi \in [0, 2\pi] & (\text{otherwise})
	\end{cases}
\end{equation}

Based on the above analysis, the proposed method for solving problem (2), termed ADMM-APG, is summarized in Algorithm 1.
\begin{algorithm}
	\caption{ADMM-APG algorithm for solving the problem (2)}
	\begin{algorithmic}
		\STATE {1: $\mathbf{Input}$:~$P, \delta_{n}^2, \mathbf{H}_1, \mathbf{H}_2, \mathbf{H}_m$,  set $\tau = 0.001, \rho = 1$, number of maximal iterations $K_{max}$;}
		\STATE {2: Initialize $\mathbf{Y}$,  $\boldsymbol{\Phi}$ and  $\mathbf{Z}$ to feasible solutions;}
         \STATE {3:        $\mathbf{for}~ k=1,2,\cdots,K_{max}$;}
		\STATE {4: Update $\mathbf{G}$ by (7);}
		\STATE {5: Update $\mathbf{Y}$ by (10);}
		\STATE {6: Update $\boldsymbol{\theta}$  by (12);}
		\STATE {7: Update $\mathbf{Z}$ by (5d);}
        \STATE {8:  $\mathbf{ End ~for}$;}
		\STATE {9: $\mathbf{Output}$:~$\mathbf{G}$ and $\boldsymbol{\theta}$.}
	\end{algorithmic}
\end{algorithm}
\subsection{Complexity and Convergence Analysis}
We assess the computational complexity (CC) of the proposed ADMM-APG algorithm by counting the number of complex multiplications (CMs). The algorithm consists of four core update procedures: the precoding matrix $\mathbf{G}$, the auxiliary matrix $\mathbf{Y}$, the phase configuration vector $\boldsymbol{\theta}$ (or matrix $\boldsymbol{\Phi}$), and the dual matrix $\mathbf{Z}$.

(1) Update of $\mathbf{G}$: Constructing the channel matrix $\mathbf{H}$ requires $M_tM_rM_i + M_rM_i$ CMs. The subsequent SVD of $\mathbf{H}$ has a complexity of $\mathcal{O}(M_tM_r\min\{M_t, M_r\})$ CMs. The overall cost for updating $\mathbf{G}$ is therefore $\mathcal{O}\left(M_tM_r\min\{M_t, M_r\} + M_tM_rM_i +M_rM_i\right)$ CMs.

(2) Update of $\mathbf{Y}$: The construction of $\mathbf{Q} = \mathbf{I}_{M_r} + C \mathbf{U} \boldsymbol{\Lambda} \boldsymbol{\Gamma} \boldsymbol{\Lambda} \mathbf{U}^\mathrm{H} + \mathbf{Z}^k$ demands $\mathcal{O}(M_rM_s^2 + \frac{1}{2}M_r^2 M_s + \frac{3}{2}M_r M_s)$ CMs. Computing $\mathbf{G}$ from Eq. (7) involves $\mathcal{O}(M_t M_s)$ CMs, while solving for $\mathbf{Y}^{k+1}$ via Eq. (25) requires $\mathcal{O}(\frac{1}{2}M_r^3 + \frac{3}{2}M_r^2)$ CMs. The dominant cost for this step is thus $\mathcal{O}(\frac{1}{2}M_r^3 + \frac{1}{2}M_r^2 M_s)$ CMs.

(3) Update of $\boldsymbol{\theta}$: The computational burden is dominated by evaluating the gradient $\nabla_{\boldsymbol{\theta}} g(\boldsymbol{\theta})$ in Eq. (14), which involves the following operations:

 Computing $\mathbf{H}\mathbf{G}^{k+1}$ requires $\mathcal{O}(M_t M_r M_s)$ CMs.

 Evaluating $\mathbf{H}\mathbf{G}^{k+1}(\mathbf{G}^{k+1})^\mathrm{H} \mathbf{H}^\mathrm{H}$ takes $\mathcal{O}(\frac{1}{2}M_r^2 M_s + \frac{1}{2}M_r M_s)$ CMs.

 Calculating $\mathbf{H}_1^\mathrm{H} \mathbf{E}$ requires $\mathcal{O}(M_r^2 M_i)$ CMs.

 Computing $\mathbf{H}_1^\mathrm{H} \mathbf{E} \mathbf{H} \mathbf{G}^{k+1}$ takes $\mathcal{O}(M_r M_i M_s)$ CMs.

 Forming $(\mathbf{G}^{k+1})^\mathrm{H} \mathbf{H}_m^\mathrm{H}$ requires $\mathcal{O}(M_t M_i M_s)$ CMs.

 Finally, evaluating $\mathrm{vec}_d \left[ \mathbf{H}_1^\mathrm{H} \mathbf{E} \mathbf{H} \mathbf{G}^{k+1} (\mathbf{G}^{k+1})^\mathrm{H} \mathbf{H}_m^\mathrm{H} \right]$ incurs $\mathcal{O}(M_s M_i)$ CMs.

The total complexity for the gradient is $\mathcal{O}(M_t M_r M_s + \frac{1}{2}M_r^2 M_s + M_r^2 M_i + M_r M_i M_s + M_t M_i M_s + M_s M_i)$ CMs. The CCs of other terms in Eq. (12) are negligible.

(4) Update of $\mathbf{Z}$: The update in Eq. (5d) requires $(M_t + 1)M_r M_i + M_t M_r M_s + \frac{1}{2}M_r^2 M_s$ CMs.

The overall complexity of the ADMM-APG algorithm, in terms of CMs, is summarized as:
\begin{equation}\label{eq16}
\begin{aligned}
\textrm{CC}_{\textrm{ADMM-APG}}=\mathcal{O}(&2M_tM_r M_i +M_tM_r\min\{M_t, M_r\}\\
&+M_{r}^{2}M_i+M_tM_iM_s+M_rM_iM_s\\
&+M_sM_i+\frac{3}{2}M_{r}^{2}M_s+ M_t M_r M_s \\
&+ \frac{1}{2}M_r^3).
\end{aligned}
\end{equation}

When the number of IRS elements  $M_i$ significantly exceeds both the number of transmit and receive antennas  $M_t$ and $M_r$, the complexity of the ADMM-APG algorithm $\textrm{CC}_{\textrm{ADMM-APG}}$ can be  approximated as $\mathcal{O}(M_tM_r M_i)$. This reflects a linear scaling with respect to  $M_i$, highlighting the algorithm's efficiency in large-IRS regimes.
\section{SIMULATION RESULTS}\label{sec:4}
In this section, we present numerical simulations to evaluate the performance of Algorithm 1 against several benchmark schemes. The network topology assumes that BS, the IRS, and receiver are positioned at the vertices of an equilateral triangle with side length $d$ meters. We adopt the Rician fading model to characterize the channels $\mathbf{H}_1$, $\mathbf{H}_2$, and $\mathbf{H}_m$, which is given by
\begin{equation}\label{17}
\mathbf{H}_{U} = \sqrt{C(d)} \left( \sqrt{\frac{\gamma}{1+\gamma}}  \mathbf{a}_r(\varphi_r) \mathbf{a}_t(\varphi_t)^H + \sqrt{\frac{1}{1+\gamma}}  \mathbf{H}_{\text{NLoS}} \right),
\end{equation}
where $C(d) = C_0(d / d_0)^{-\beta}$ represents the path loss at distance $d$, with $C_0$ being the reference path loss at $d_0 = 1~\text{m}$ and $\beta$ the path-loss exponent. The Rician factor is denoted by $\gamma$, while $\varphi_t$ and $\varphi_r \in [0, 2\pi)$ correspond to the azimuth angles of departure and arrival of the line-of-sight (LoS) component, respectively. The array response vectors at the transmitter and receiver, $\mathbf{a}_t(\varphi)$ and $\mathbf{a}_r(\varphi)$, are modeled under a uniform linear array configuration with $N$ elements. Specifically,
\begin{equation}\label{18}
\mathbf{a}(\varphi) = \frac{1}{\sqrt{N}} \left[1,\ e^{j m d_{as} \sin \varphi},\ \ldots,\ e^{j m d_{as} (N-1) \sin \varphi} \right]^T,
\end{equation}
where $m = 2\pi / \lambda$, $\lambda$ is the carrier wavelength, and $d_{as}$ is the antenna spacing. The non-line-of-sight (NLoS) component $\mathbf{H}_{\text{NLoS}}$ consists of independent and identically distributed entries drawn from a complex Gaussian distribution with zero mean and unit variance.

The system under consideration employs uniform linear arrays (ULAs) at both the transmitter and receiver, with $M_t=16$ and $M_r=4$ antennas, respectively. The number of data streams is set to $M_s=4$, and an IRS comprising $M_i=100$ elements is assumed. The simulation parameters are as follows, unless otherwise specified: the reference path loss $C_0 = -30~\text{dB}$, the distance $d = 30~\text{m}$, the noise power $\sigma_n^2 = 1$, the antenna spacing $d_{as} = \lambda/2$, the Rician factor $\gamma = 10~\text{dB}$, and the path-loss exponent $\beta = 2$. The number of maximal iterations $K_{max}=100$. All presented results are averaged over 1,000 independent channel realizations.
\subsection{Spectral efficiency versus transmit power}
Fig.~2 compares the spectral efficiency of various methods against transmit power. The ADMM-APG algorithm demonstrates a significant advantage over all benchmark algorithms across the entire power range. While its performance is comparable to LADMM at -10 dB, the superiority of ADMM-APG becomes increasingly pronounced at higher power levels. Notably, in the medium-to-high power region, ADMM-APG exhibits a substantially faster growth rate. At 20 dB, it achieves a spectral efficiency of 68 bps/Hz, outperforming LADMM, AO, PGM, and SPGM by 4, 7, 8, and 15 bps/Hz, respectively.  Furthermore, schemes with IRS phase optimization yield higher spectral efficiency than their non-optimized counterparts (without IRS or random phase), underscoring the critical importance of phase optimization for IRS-assisted MIMO systems.
\begin{figure}
	\centering
	\includegraphics[width=3in,angle=0]{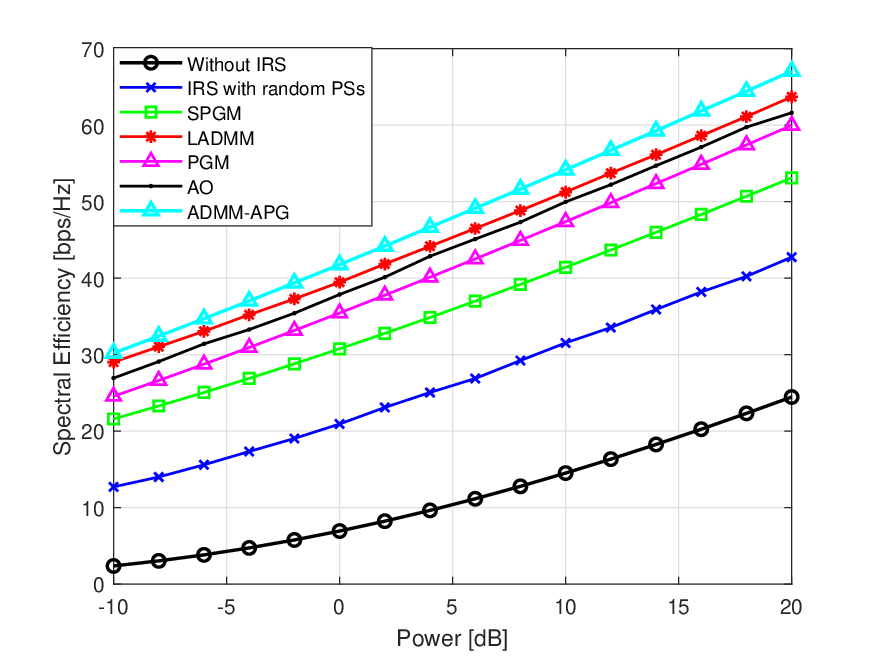}
	\caption{Spectral efficiency under different transmit power }
	\label{fig2}
\end{figure}
\begin{figure}
	\centering
	\includegraphics[width=3in,angle=0]{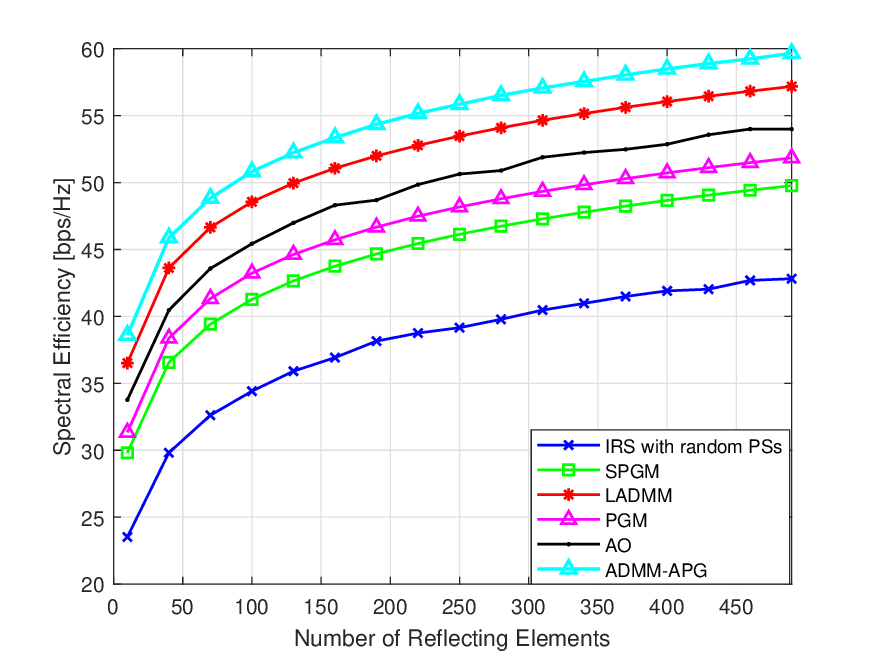}
	\caption{Spectral efficiency versus number of reflecting elements}
	\label{fig3}
\end{figure}
\begin{figure}
	\centering
	\includegraphics[width=3in,angle=0]{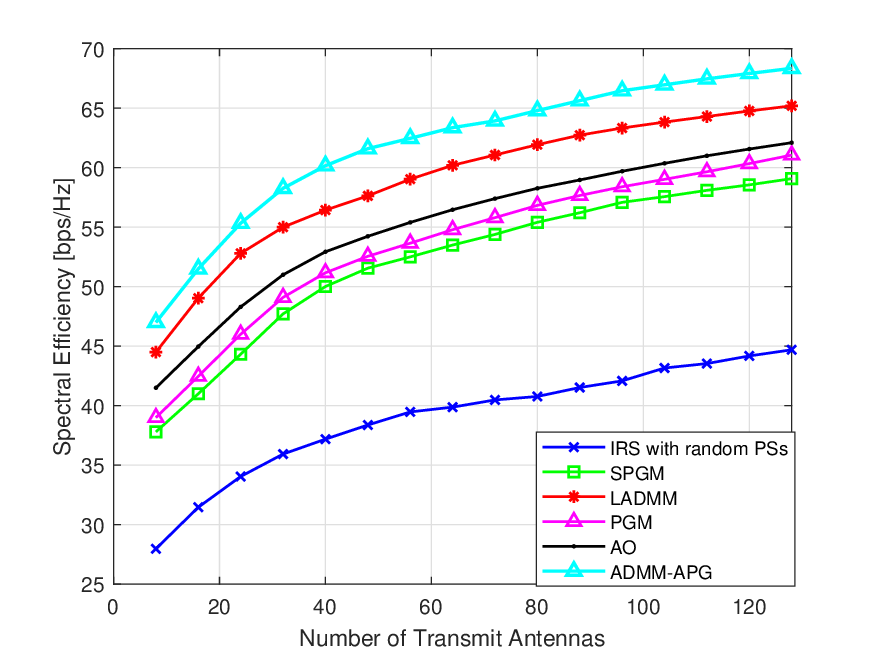}
	\caption{Spectral efficiency versus number of transmit antennas }
	\label{fig4}
\end{figure}
\begin{figure}
	\centering
	\includegraphics[width=3in,angle=0]{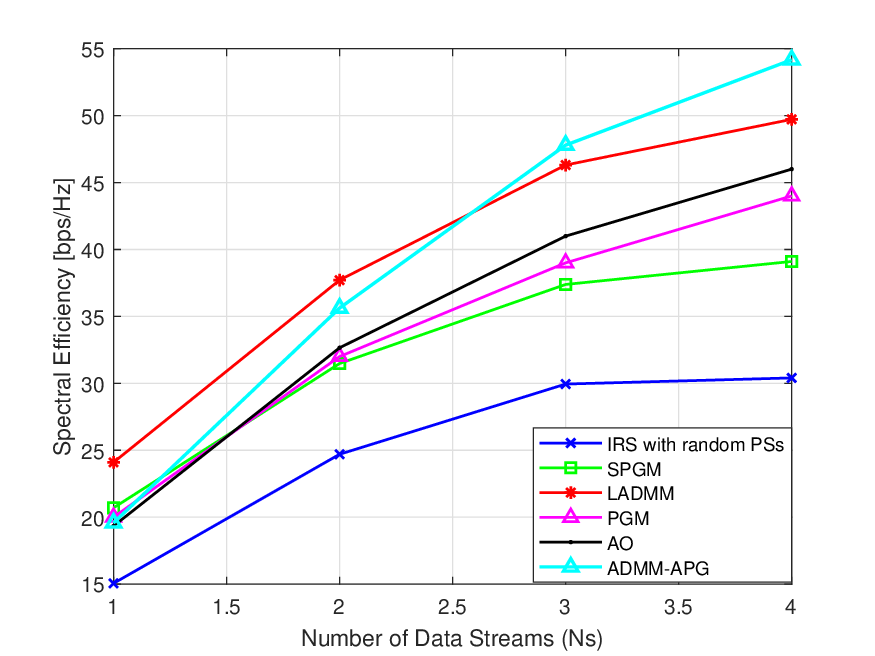}
	\caption{Spectral efficiency versus number of data streams }
	\label{fig5}
\end{figure}
\subsection{Spectral efficiency versus number of IRS reflecting elements}
 A key observation from Fig. 3, which plots spectral efficiency against the number of IRS elements $M_i$ from 10 to 500 at $10$ dB transmit power, is the consistent outperformance of the proposed ADMM-APG algorithm. At $M_i=10$, ADMM-APG achieves $38$ bps/Hz, marking a  $2$ bps/Hz gain over the  $36$ bps/Hz attained by LADMM, a result aligning with Fig. 2. This advantage persists, averaging $3$ bps/Hz throughout the evaluated range. Furthermore, the substantial superiority of phase-optimized schemes over the random phase baseline highlights the necessity of phase shift design for effective IRS-assisted MIMO communications.
 \subsection{Spectral efficiency versus number of transmit antennas}
As shown in Fig. 4, the spectral efficiency of the proposed ADMM-APG algorithm is evaluated against an increasing number of transmit antennas (from 8 to 128) at a transmit power of  $10$ dB. The ADMM-APG algorithm not only demonstrates robust performance superiority across all scales but also exhibits significantly steeper growth, decisively outperforming other methods when the antenna number is large. This compelling performance advantage, coupled with its lower computational complexity, as will be verified in Table I, establishes the practical value of the ADMM-APG approach.
\subsection{Spectral efficiency versus number of data streams}
Fig. 5 presents a comparison of spectral efficiency as a function of the number of data streams, $M_s$. It is observed that for $M_s=1$, the LADMM algorithm yields the highest spectral efficiency, with the proposed ADMM-APG method performing similarly to PGM, AO, and SPGM. With an increasing number of data streams, all algorithms exhibit improved performance, but the ADMM-APG algorithm demonstrates a markedly steeper growth rate. Notably, ADMM-APG matches the performance of LADMM at $M_s=2.5$ and obtains a notable gain of $5$ bps/Hz over it at $M_s=4$.
\subsection{Spectral efficiency versus channel estimation error}
To evaluate the robustness of different optimization methods to channel estimation errors in IRS-assisted MIMO systems, Fig.~6 plots the spectral efficiency against the channel estimation error. The imperfect channel estimate is modeled as [28], [39]:

\begin{equation}\label{eq19}
\bar{\mathbf{H}}=\mathbf{H}+\mathbf{\triangle H},
\end{equation}
where $\mathbf{H}$ is the true channel, and the estimation error $\mathbf{\triangle H}\sim \mathcal{CN}(0, \gamma_{H}^{2}\mathbf{I})$ is characterized by $\gamma_{H}^{2}=\delta||\mathbf{H}||^2/\sqrt{M_tM_r}$.

As observed in Fig. 6, spectral efficiency degrades with increasing channel estimation error $\delta$ across all methods. The proposed ADMM-APG algorithm, however, demonstrates significantly stronger robustness to CSI imperfections compared to alternatives. At $\delta=0.4$, ADMM-APG incurs only a $18.0\%$ performance degradation versus $29.8\%$ for LADMM; this trend continues at $\delta=0.9$, with corresponding losses of $19.3\%$ and $33.7\%$. Notably, these relative advantages coincide with substantially higher absolute performance: ADMM-APG achieves spectral efficiencies $21.4\%$ and $31.7\%$ greater than LADMM at $\delta=0.4$ and $\delta=0.9$, respectively. The proposed method also consistently surpasses other benchmarks (PGM, AO, SPGM) under all tested error conditions. Finally, the noticeable gap between all optimization-based methods and the random phase baseline reaffirms the essential role of deliberate phase design, even with imperfect CSI.
\subsection{Computational complexity comparison}
\begin{table*}[t]
	\centering
	\caption{Computational complexity comparison among various methods}
	\begin{tabular}{|c|c|c|c|}
		\hline
		Method & Computational Complexity $\mathcal{O}\left ( \cdot  \right)$ &Iteration Number& Total CC \\
		\hline
		ADMM-APG &
		$\begin{aligned}
      &2M_tM_r M_i + M_tM_r\min\{M_t, M_r\} + M_{r}^{2}M_i\\
      & + M_tM_iM_s + M_rM_iM_s + M_sM_i + \frac{3}{2}M_{r}^{2}M_s\\
      & + M_t M_r M_s + \frac{1}{2}M_r^3.
      \end{aligned}$ & 10& 234400 \\
		\hline
		LADMM &
$\begin{aligned}
&M_tM_r\min\{M_t, M_r\} + M_{i}^{2}M_t+M_tM_rM_i+I_{L}M_{i}^{2}.\\
& I_L ~\text{is the number of iterations}
\end{aligned}$  & 20& 366656\\
		\hline
		AO &
$\begin{aligned}
    &((L_{AO}+1)M_tM_r M_i + L_{AO}(r^{3}+ \tfrac{1}{2}M_{t}^{2}r) \\
    &+ I_{AO}[M_{t}^{3}+M_{t}^{2}M_i+2M_tM_r M_i \\
    &+(2M_tM_{r}^{2}+2M_{r}^{3})M_i+r^3+\tfrac{1}{2}M_{t}^{2}r]), \\
    &\text{where}   ~ L_{AO} ~\text{is the number of independent realizations}\\
    & \text{ of}~\theta_n, n=1,2,\cdots,M_i, r=\min\{M_t, M_r\}.
\end{aligned}$  & 10& 1774720\\
		\hline
		PGM &
		$\begin{aligned}
    &2M_tM_r M_i + 2M_{t}^{2}M_r + M_{r}M_i + M_tM_i \\
    &+ \frac{3}{2}M_{t}^{3}+\frac{3}{2}M_{r}^{2}M_t + 3M_i +M_r^3.
    \end{aligned}$ & 10& 237400\\
		\hline
		SPGM &
$\begin{aligned}
&M_tM_r\min\{M_t, M_r\} + M_{i}^{2}M_t+M_tM_rM_i+I_{s}M_{i}^{3}.\\
& I_s ~\text{is the number of iterations}
\end{aligned}$  & 20& 20166656\\
		\hline
	\end{tabular}
	\label{tab:complexity_comparison}
\end{table*}
\begin{figure}
	\centering
	\includegraphics[width=3in,angle=0]{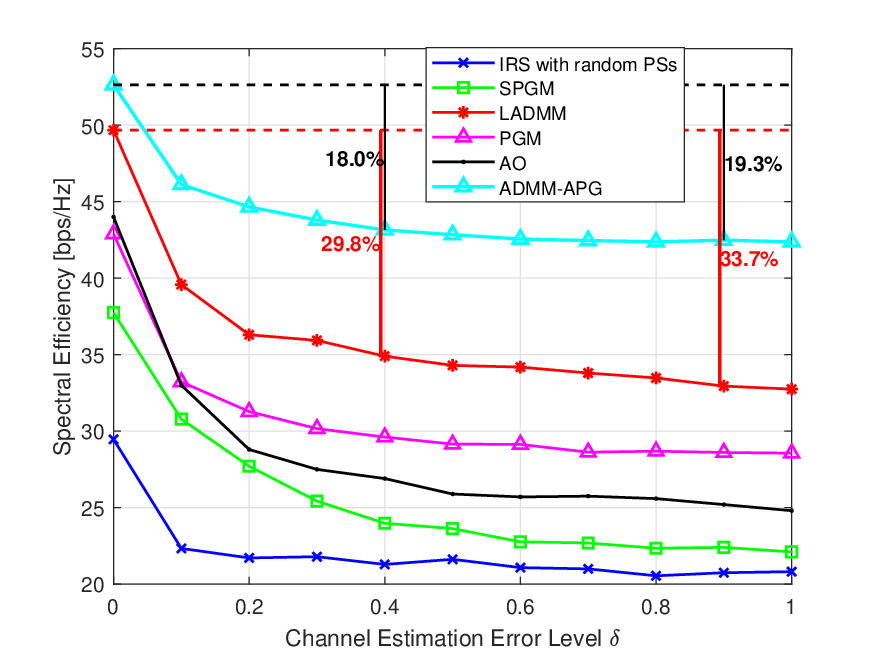}
	\caption{Spectral efficiency under channel estimation errors}
	\label{fig6}
\end{figure}

\begin{figure}
	\centering
	\includegraphics[width=3in,angle=0]{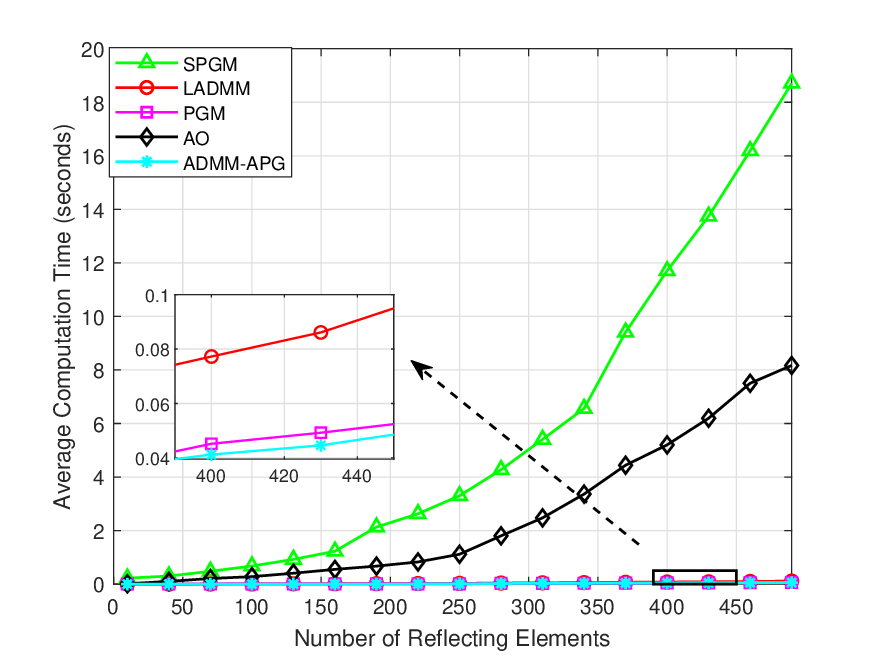}
	\caption{Average compute time versus IRS Elements}
	\label{fig7}
\end{figure}

\begin{figure}
	\centering
	\includegraphics[width=3in,angle=0]{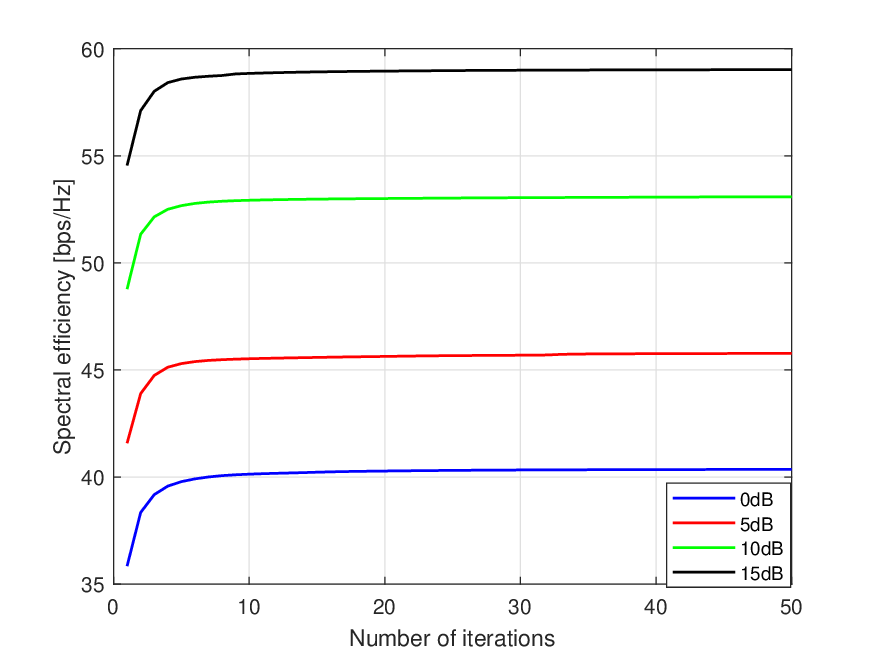}
	\caption{Spectral efficiency of ADMM-APG under different power levels}
	\label{fig8}
\end{figure}

A comparative analysis of computational complexity across different methods is presented in Table I. Among the convergent algorithms, SPGM exhibits the highest computational burden due to its cubic complexity order $\mathcal{O}(M_i^3)$, while ADMM-APG achieves the most efficient implementation. Although ADMM-APG, LADMM, and PGM share the same order of magnitude ($10^5$), ADMM-APG delivers superior spectral efficiency across all evaluation metrics, as substantiated by Figs.~2-6. Quantitatively, ADMM-APG requires only 63.93\% and 98.74\% of the computational load of LADMM and PGM, respectively.

While ADMM-APG and PGM demonstrate comparable complexity in conventional MIMO configurations, their scalability diverges significantly in massive MIMO regimes. The computational demand of PGM grows substantially faster due to its dominant $M_{t}^{3}$ term-a component absent in ADMM-APG's complexity structure. For instance, in an IRS-assisted massive MIMO setup with $M_t=64$, $M_r=8$, and $M_i=600$, ADMM-APG and PGM require 834,784 and 1,063,368 complex multiplications, respectively. This divergence stems from their distinct complexity dependencies: ADMM-APG scales with $2M_tM_rM_i$, whereas PGM depends on $2M_tM_rM_i + \frac{3}{2}M_t^3$.

Consequently, ADMM-APG emerges as the preferable candidate for spectral efficiency maximization in both conventional and massive MIMO systems. This conclusion is further reinforced by Fig.~7, which compares average computation time for $M_t=16$, $M_r=M_s=4$. The observed timing results align consistently with the complexity analysis in Table I, validating our theoretical assessment.
\subsection{Spectral efficiency versus number of iterations}
Fig. 8 illustrates the convergence characteristics of the proposed ADMM-APG algorithm across different transmit power levels (0, 5, 10, and 15 dB). The algorithm demonstrates consistently rapid and stable convergence in all power configurations, typically achieving spectral efficiency stabilization within approximately 10 iterations significantly fewer than the maximum iteration limit of 100.
~At 10 dB transmit power, the algorithm attains a converged spectral efficiency of 53 bps/Hz, which corresponds closely with the performance levels observed in Fig. 2. This consistency across experimental configurations further validates the reliability of the proposed ADMM-APG method.

\section{CONCLUSION}
This paper addresses the spectral efficiency maximization problem in IRS-assisted MIMO communication systems. We propose a novel ADMM-APG algorithm that jointly optimizes the transmit precoding matrix and the IRS phase shift matrix. Under the ADMM framework, the original problem is decomposed into tractable subproblems corresponding to the precoding matrix, an auxiliary matrix, and the phase shift matrix. To handle the non-convex unit-modulus constraints on the phase shifts, the APG method is incorporated into the ADMM procedure. Each resulting subproblem admits a closed-form solution, enabling efficient implementation of the overall algorithm while substantially improving convergence speed and reducing computational complexity.
Simulation results show that the proposed ADMM-APG algorithm comprehensively outperforms existing mainstream methods in terms of achievable spectral efficiency. Moreover, under imperfect channel state information, the proposed approach demonstrates superior robustness, with performance degradation significantly milder than that of benchmark algorithms. These findings collectively validate the effectiveness and practical advantages of the ADMM-APG method for spectral efficiency maximization in IRS-aided MIMO systems.

\begin{appendices}
	\section{ Derivation of formula (10) }
To solve this subproblem, we take the derivative of the augmented Lagrangian function with respect to $\mathbf{Y}$ and set the resulting expression equal to zero, yielding:
\begin{equation}
	\begin{aligned}
		&-\mathbf{Y}^{-1} + \rho\left( \mathbf{Y} - \mathbf{I}_{M_r} - C \mathbf{U} \boldsymbol{\Lambda} \boldsymbol{\Gamma} \boldsymbol{\Lambda} \mathbf{U}^\mathrm{H} + \mathbf{Z}^k \right) = \mathbf{0} \\
		&-\mathbf{Y}^{-1} + \rho \mathbf{Y} = \rho\left( \mathbf{I}_{M_r} + C \mathbf{U} \boldsymbol{\Lambda} \boldsymbol{\Gamma} \boldsymbol{\Lambda} \mathbf{U}^\mathrm{H} - \mathbf{Z}^k \right) \\
		&\mathbf{Y} - \rho^{-1} \mathbf{Y}^{-1} = \mathbf{I}_{M_r} + C \mathbf{U} \boldsymbol{\Lambda} \boldsymbol{\Gamma} \boldsymbol{\Lambda} \mathbf{U}^\mathrm{H} - \mathbf{Z}^k
	\end{aligned}
\end{equation}

This equation belongs to a nonlinear matrix equation, making direct solution challenging. To solve it efficiently, we employ an eigenvalue decomposition approach, transforming the problem into solving a system of scalar equations. Specifically, we first perform eigenvalue decomposition on the right-hand matrix $\mathbf{Q}=\mathbf{I}_{M_r} + C \mathbf{U} \boldsymbol{\Lambda} \boldsymbol{\Gamma} \boldsymbol{\Lambda} \mathbf{U}^\mathrm{H} + \mathbf{Z}^k$, thus we have  $\mathbf{Q}=\mathbf{U}_1 \boldsymbol{\Lambda}_1 \mathbf{U}_1^H$, where $\mathbf{U}_1$ is a unitary matrix,  and $\boldsymbol{\Lambda}_1$ is a diagonal matrix with diagonal elements satisfying $\boldsymbol{\Lambda}_1 = \mathrm{diag}(\lambda_1, \dots, \lambda_{M_r}), \quad \lambda_i > 0$. By left-multiplying Eq.~(20) by $\mathbf{U}_1^H$ and right-multiplying it by $\mathbf{U}_1$, we obtain:
\begin{equation}
	\mathbf{U}_1^H \left( \mathbf{Y} - \rho^{-1} \mathbf{Y}^{-1} \right) \mathbf{U}_1 = \boldsymbol{\Lambda}_1
\end{equation}	
	
Let $ \widetilde{\mathbf{Y}} = \mathbf{U}_1^H \mathbf{Y} \mathbf{U}_1$,  we can obtain $\widetilde{\mathbf{Y}}^{-1} = \mathbf{U}_1^H \mathbf{Y}^{-1} \mathbf{U}_1$. Therefore,  equation (21) is simplified  to:
\begin{equation}
	\widetilde{\mathbf{Y}} -\rho^{-1} \widetilde{\mathbf{Y}}^{-1} = \boldsymbol{\Lambda}_1
\end{equation}

Since $\boldsymbol{\Lambda}_1$ is a diagonal matrix, we can derive a diagonal solution for the above equation by solving:
\begin{equation}
	\widetilde{{Y}}_{ii} - \rho^{-1} \widetilde{{Y}}_{ii}^{-1} = \lambda_i, \quad i = 1, 2, \dots, M_r
\end{equation}

Multiplying both sides by $\widetilde{{Y}}_{ii}$ yields a quadratic equation in terms of $\widetilde{{Y}}_{ii}$ :
\begin{equation}
	\widetilde{{Y}}_{ii}^2 - \lambda_i \widetilde{{Y}}_{ii} - \rho^{-1} = {0}
\end{equation}

The solution is $\widetilde{{Y}}_{ii} = \frac{\lambda_i + \sqrt{\lambda_i^2 + 4\rho^{-1}}}{2}$, where $\widetilde{{Y}}_{ii}>0$. Thus, $\widetilde{\mathbf{Y}} = \mathrm{diag}\left( \widetilde{{Y}}_{11}, \widetilde{{Y}}_{22}, \dots, \widetilde{{Y}}_{M_rM_r} \right)\succ0$ is a positive matrix.
Since $ \widetilde{\mathbf{Y}} = \mathbf{U}_1^H \mathbf{Y} \mathbf{U}_1$, the final closed-form solution in terms of $\mathbf{Y}$ is given by:
\begin{equation}
	\mathbf{Y}^{k+1} = \mathbf{U}_1 \widetilde{\mathbf{Y}} \mathbf{U}_1^H
\end{equation}	
	\section{ Derivation of formula (14) }
Based on the notation established in \cite{IEEEconf:48}, the complex differential of the function \( g(\boldsymbol{\theta}) \)
in  (14) is denoted as \( \mathrm{d}g(\boldsymbol{\theta}) \).
~Let		$\mathbf{E} = \mathbf{Y}^{k+1} - \mathbf{I}_{M_r} - C \mathbf{H} \mathbf{G}^{k+1} (\mathbf{G}^{k+1})^H \mathbf{H}^\mathrm{H} + \mathbf{Z}^k$, we have
\begin{equation}
	\begin{aligned}
		\mathrm{d}g &= 2\,\mathrm{Tr}\bigl[\mathbf{E}\,\mathrm{d}(\mathbf{E})\bigr] \\
		&= -2C\,\mathrm{Tr} \Bigl\{ \mathbf{E} \Bigl[ \mathrm{d}(\mathbf{H})\,\mathbf{G}^{K+1}\,(\mathbf{G}^{k+1})^H\,\mathbf{H}^\mathrm{H}\\
 &~~~~+ \mathbf{H}\,\mathbf{G}^{K+1}\,(\mathbf{G}^{k+1})^H\,\mathrm{d}(\mathbf{H}^\mathrm{H}) \Bigr] \Bigr\} \\
		&= -2C\,\mathrm{Tr} \Bigl\{ \mathbf{E}\Bigl[ \mathbf{H}_1\,\mathrm{d}(\boldsymbol{\Theta})\,\mathbf{H}_m\,\mathbf{G}^{K+1}\,(\mathbf{G}^{k+1})^H\,\mathbf{H}^\mathrm{H}\\
 &~~~~+ \mathbf{H}\,\mathbf{G}^{K+1}\,(\mathbf{G}^{k+1})^H\,\mathbf{H}_{m}^{H}\,\mathrm{d}(\boldsymbol{\Theta})^\mathrm{H}\,\mathbf{H}_1^\mathrm{H} \Bigr] \Bigr\} \\
		&\overset{(a)}{=} -2C\,\mathrm{Tr} \Bigl[ \mathbf{H}_m\,\mathbf{G}^{K+1}\,(\mathbf{G}^{k+1})^H\,\mathbf{H}^\mathrm{H}\,\mathbf{E}\,\mathbf{H}_1\,\mathrm{d}(\boldsymbol{\Theta})^\mathrm{H} \\
&~~~~ + \mathbf{H}_1^\mathrm{H}\,\mathbf{E}\,\mathbf{H}\,\mathbf{G}^{K+1}\,(\mathbf{G}^{k+1})^H\,\mathbf{H}_{m}^{H}\,\mathrm{d}(\boldsymbol{\Theta})^\mathrm{H} \Bigr] \\
		&\overset{(b)}{=} -2C\,\mathrm{Tr} \Bigl[ \mathbf{H}_m\,\mathbf{G}^{K+1}\,(\mathbf{G}^{k+1})^H\,\mathbf{H}^\mathrm{H}\,\mathbf{E}\,\mathbf{H}_1\,\mathrm{d}(\boldsymbol{\Theta})^\mathrm{H} \\
&~~~~+ \mathbf{H}_{m}^{*}\,(\mathbf{G}^{K+1})^{*}\,(\mathbf{G}^{K+1})^\mathrm{T}\,\mathbf{H}^\mathrm{T}\,\mathbf{E}^\mathrm{T}\,\mathbf{H}_1^*\,\mathrm{d}(\boldsymbol{\Theta})^{*} \Bigr] \\
		&\overset{(c)}{=} -2C \Bigl\{ \mathrm{vec}^\mathrm{T} \Bigl[ \bigl( \mathbf{H}_m\,\mathbf{G}^{K+1}\,(\mathbf{G}^{K+1})^\mathrm{H}\,\mathbf{H}^\mathrm{H}\,\mathbf{E}\,\mathbf{H}_1 \bigr)^\mathrm{T} \Bigr]\\
&~~~~\times \textbf{L}_d\,\mathrm{d}\bigl(\mathrm{vec}(\boldsymbol{\Theta})\bigr)\\
 &~~~~+ \mathrm{vec}^\mathrm{T} \Bigl[ \bigl( \mathbf{H}_{m}^{*}\,(\mathbf{G}^{K+1})^{*}\,(\mathbf{G}^{K+1})^\mathrm{T}\,\mathbf{H}^\mathrm{T}\,\mathbf{H}_1^* \bigr)^\mathrm{T}\,\\
 &~~~~\times \textbf{L}_d\,\mathrm{d}\bigl(\mathrm{vec}(\boldsymbol{\Theta})^*\bigr) \Bigr] \Bigr\}
	\end{aligned}
\end{equation}

Building on the trace identity $\mathrm{Tr}(\mathbf{A}\mathbf{B}) = \mathrm{Tr}(\mathbf{B}\mathbf{A})$, step $(a)$ follows directly. Step $(b)$ applies the property $\mathrm{Tr}(\mathbf{A}^{H}\mathbf{B}^{H}) = \mathrm{Tr}(\mathbf{A}^{*}\mathbf{B}^{*})$, while $(c)$ employs the relation $\mathrm{Tr}(\mathbf{A}^{T}\mathbf{B}) = \textrm{vec}^{T}(\mathbf{A})\textrm{vec}(\mathbf{B})$. Here, $\textbf{L}_d$ denotes a placement matrix that maps the diagonal elements of a square matrix $\mathbf{A}$ to corresponding positions in $\textrm{vec}(\mathbf{A})$.

The gradient of $g(\boldsymbol{\theta})$ is subsequently derived from expression (26).
	\begin{equation}
		\begin{aligned}
			\nabla g(\boldsymbol{\theta}) &= -2C\textbf{L}_d^\mathrm{T}\, \mathrm{vec} \left[ \left( \mathbf{H}_m \mathbf{G}^{k+1} (\mathbf{G}^{k+1})^\mathrm{H} \mathbf{H}^\mathrm{H} \mathbf{E} \mathbf{H}_1 \right)^\mathrm{T} \right] \\
			&\overset{(d)}{=}-2C\mathrm{vec}_{d} \left( \mathbf{H}_1^\mathrm{H} \mathbf{E}^\mathrm{H} \mathbf{H} \mathbf{G}^{k+1} (\mathbf{G}^{k+1})^\mathrm{H} \mathbf{H}_m^\mathrm{H} \right) \\
			&= -2C\mathrm{vec}_{d} \left( \mathbf{H}_1^\mathrm{H} \mathbf{E} \mathbf{H} \mathbf{G}^{k+1} (\mathbf{G}^{k+1})^\mathrm{H} \mathbf{H}_m^\mathrm{H} \right) \\
			&= -2C\mathrm{vec}_{d} \Bigl[ \mathbf{H}_1^\mathrm{H} \mathbf{E} \left( \mathbf{H}_1 \boldsymbol{\Theta} \mathbf{H}_m + \mathbf{H}_2 \right) \mathbf{G}^{k+1} \\ 
            &~~~~\times (\mathbf{G}^{k+1})^\mathrm{H} \mathbf{H}_m^\mathrm{H} \Bigr]
		\end{aligned}
	\end{equation}
where $(d)$ employs the property $\textbf{L}_d^\mathrm{T}\mathrm{vec}(\mathbf{A})=\mathrm{vec}_{d}(\mathbf{A})$.
\end{appendices}

\end{document}